\documentclass[a4paper,11pt]{article}
\usepackage{pos}
\usepackage{lineno}
\usepackage{hyperref}
\usepackage{url}

\title{A new view of UHECRs with the Pierre Auger Observatory}

\author{Denise Boncioli$^{a}$\footnote{Speaker} for the Pierre Auger Collaboration$^{b}$}

\affiliation[a]{Universit\`{a} degli Studi dell'Aquila, Dipartimento di Scienze fisiche e chimiche, via Vetoio, L'Aquila, Italy}

\affiliation[b]{Observatorio Pierre Auger, Av. San Mart\'{i}n Norte 304, 5613 Malarg\"{u}e, Argentina\\
Full author list: \texttt{\url{https://www.auger.org/archive/authors\_2024\_11.html}
}}

\emailAdd{spokespersons@auger.org}

\abstract{In its Phase I, the Pierre Auger Observatory has led to several observations, driving the field of ultra-high-energy cosmic ray (UHECR) research over the last 20 years. Major achievements obtained so far include the unprecedented precise energy spectrum and its features, the observables linked to the UHECR mass composition and the distribution of arrival directions of the most energetic events. These results, together with the non-observation of high-energy neutrinos and photons, strongly disfavor the pre-Auger pure-proton paradigm. 

In this talk, we will provide an overview on the main results of the Observatory, and describe possible astrophysical scenarios for their interpretation. The prospects of improving the current understanding about UHECR characteristics during the Phase II of the Observatory will be also shown.}

\FullConference{7th International Symposium on Ultra High Energy Cosmic Rays (UHECR2024)\\
 17-21 November 2024\\
Malarg\"{u}e, Mendoza, Argentina\\}

\begin{document}
\maketitle
\section{Introduction}
\vspace{-0.3cm}
The Pierre Auger Observatory is the largest observatory for measuring ultra-high-energy comic rays (UHECRs) ever built. Its construction started in 1999, the detector deployment started in 2002 and was completed in 2008. It has recently entered Phase II, which is expected to run until 2035. 

In its first 20 years of operation, the Pierre Auger Observatory has measured UHECRs revolutionizing old paradigms about their nature and revealing unexpected scenarios. 
While providing the most precise measurements  necessary for the understanding of UHECR characteristics, the Pierre Auger Observatory provides us with insights on particle physics details. Physics beyond standard model can be investigated with measurements at the highest energies, mostly thanks to the sensitivity to cosmogenic particles and to the fraction of protons in UHECRs.

In this contribution, the main experimental results will be reviewed, and the picture of UHECRs emerging from them will be discussed, taking into account the current precision and complexity of the data collected in Phase I, and the expected achievements of Phase II.

\section{The Pierre Auger Observatory and the AugerPrime upgrade}
\vspace{-0.3cm}
\begin{figure}[t]
\hspace{0.7cm}
\begin{minipage}{5cm}
\includegraphics[scale=0.35]{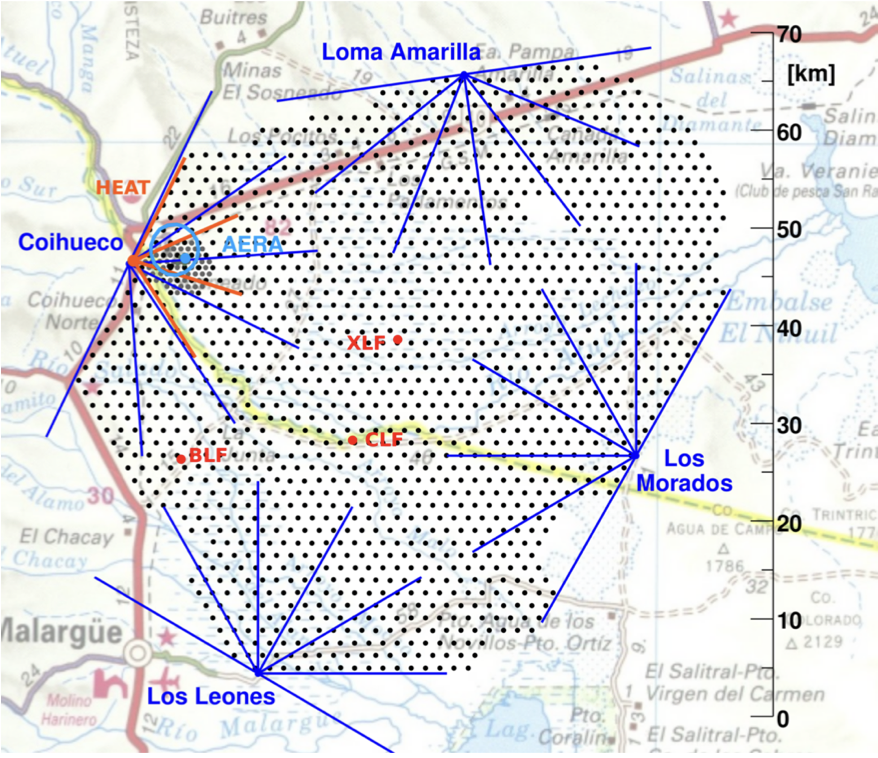}
\end{minipage}
\hspace{2.9cm}
\begin{minipage}{5cm}
\includegraphics[scale=0.25]{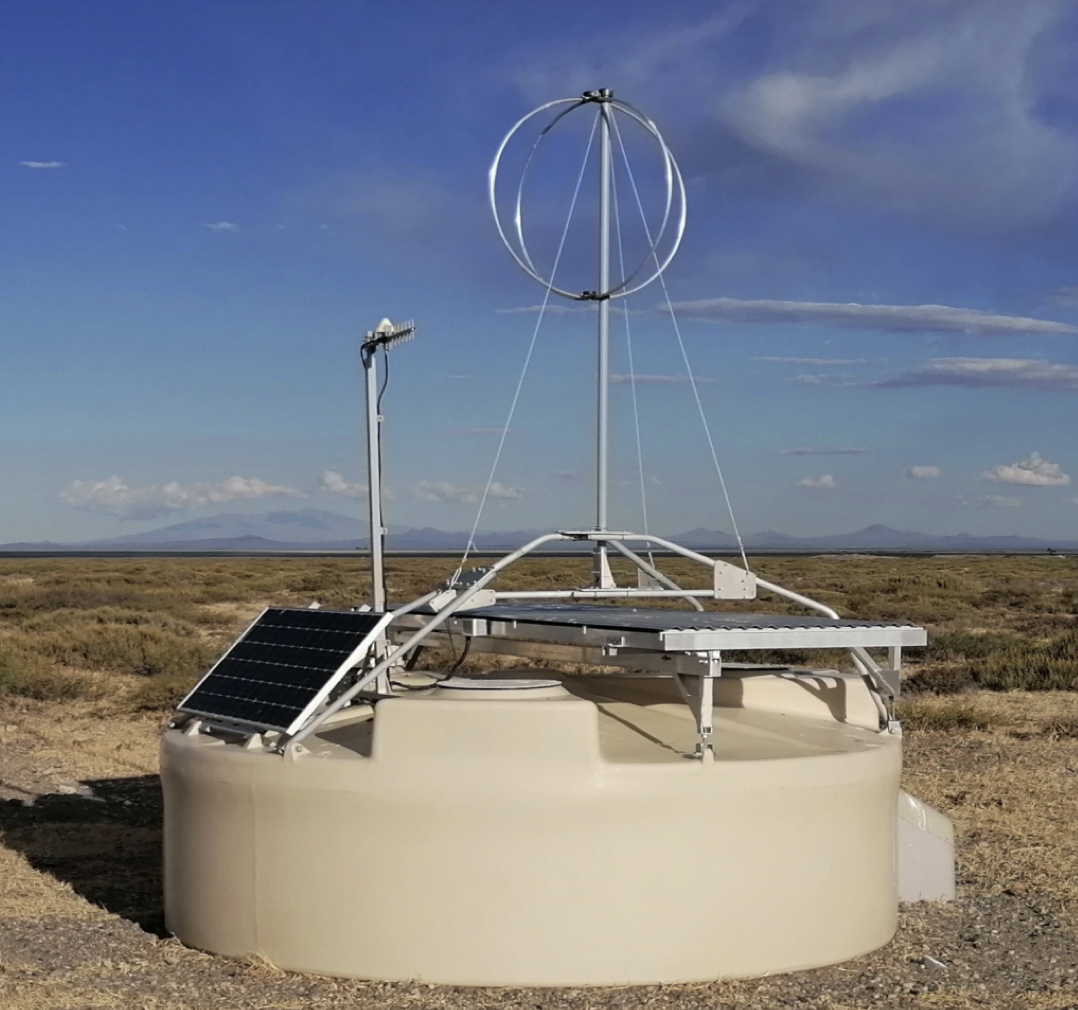}
\end{minipage}
\caption{Left: layout of the Pierre Auger Observatory, with the WCD (black dots) and the azimuthal field of view of the FD telescopes (blue and red lines). The laser facilities for the atmospheric monitoring are indicated with red dots, the AERA detectors in cyan. Right: a WCD station equipped with AugerPrime detectors (SSD positioned on top of the WCD, and the radio detector on top of the SSD).}
\label{map}
\end{figure}
The Pierre Auger Observatory~\cite{PierreAuger:2015eyc}, whose layout is shown schematically in Fig.~\ref{map} (left), is located near Malarg\"{u}e, in Argentina. It consists of a surface detector (SD) array of about 1600 water Cherenkov detectors (WCD), arranged in a 1.5 km grid over a total area of about 3000 km$^{2}$ (see Fig.~\ref{map}, right). They measure the density of particles in the extensive air shower (EAS) as it reaches the ground, and the distribution of their arrival times. The fluorescence detector (FD) consists of 24 telescopes that span over 3$^{\circ}$ to 30$^{\circ}$ of elevation and are grouped in 4 sites. They measure the energy deposited in the atmosphere through the fluorescence light produced during the development of the EAS. Two additional denser arrays of WCDs are deployed with detector spacing of 750 m and 433 m, together with three high-elevation fluorescence telescopes (HEAT), with elevation angle from 30$^{\circ}$ to 60$^{\circ}$, needed to extend the energy range down to about 10$^{16.5}$ eV. The Auger Engineering Radio Array (AERA), with more than 150 radio antennas, measures low energy showers by detecting their radio emission. Several devices are employed to monitor the atmosphere. 

Several upgrades have been implemented and the Phase II of the Pierre Auger Observatory has started. Scintillator surface detectors (SSD) are installed above the existing WCDs (see Fig.~\ref{map}, right panel), in order to improve the discrimination of the electromagnetic and muonic components of the showers detected up to zenith angles of about 60$^{\circ}$. Small photomultipliers (sPMT) are installed together with the existing three large PMTs at each WCD, to reduce the occurrence of saturated signals in the WCDs closest to the shower axis. Radio detectors (RD) are mounted above each of WCD with the aim of adding composition information for showers detected at zenith angles larger than 65$^{\circ}$. Underground muon detectors (UMD) complement the information of the muon content of the shower and its time structure. New electronics (upgraded unified board, UUB) will process the WCD, SSD and RD signals, and provide the trigger for the UMD. A more detailed description of the AugerPrime detectors, the expected data and scientific reach can be found in \cite{David:UHECR24}.

\section{A new view of UHECRs with the Pierre Auger Observatory}
\vspace{-0.3cm}
\begin{figure}[t]
\hspace{-0.8cm}
\begin{minipage}{5cm}
\includegraphics[scale=0.39]{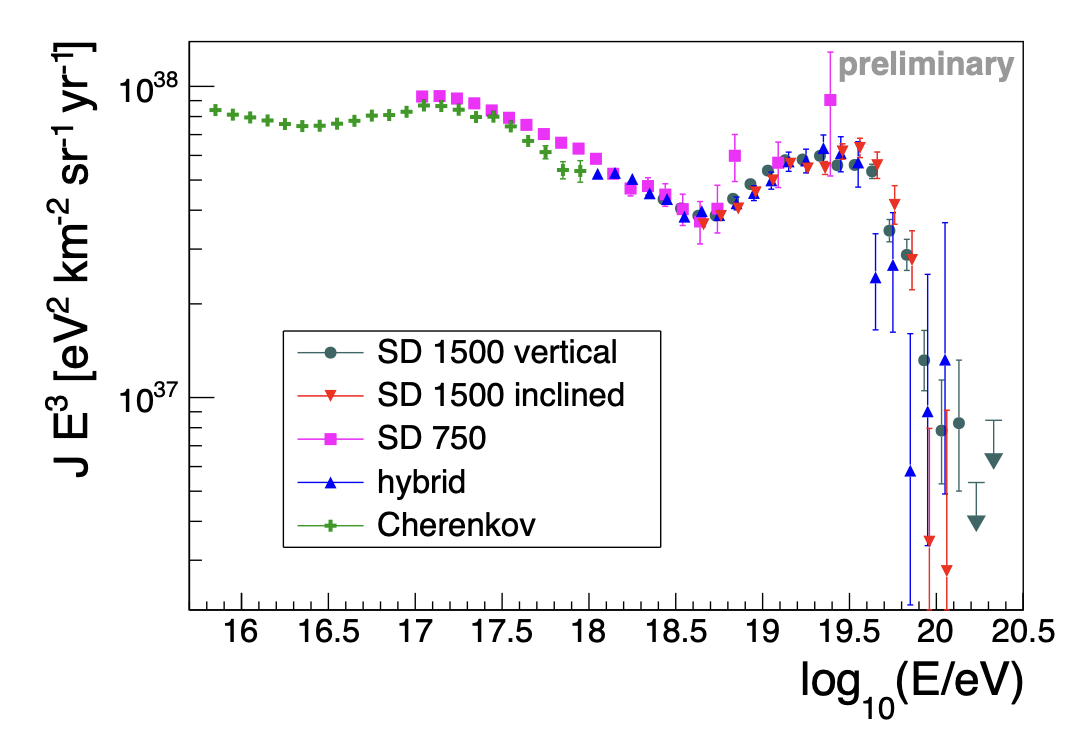}
\end{minipage}
\hspace{2.7cm}
\begin{minipage}{5cm}
\includegraphics[scale=0.32]{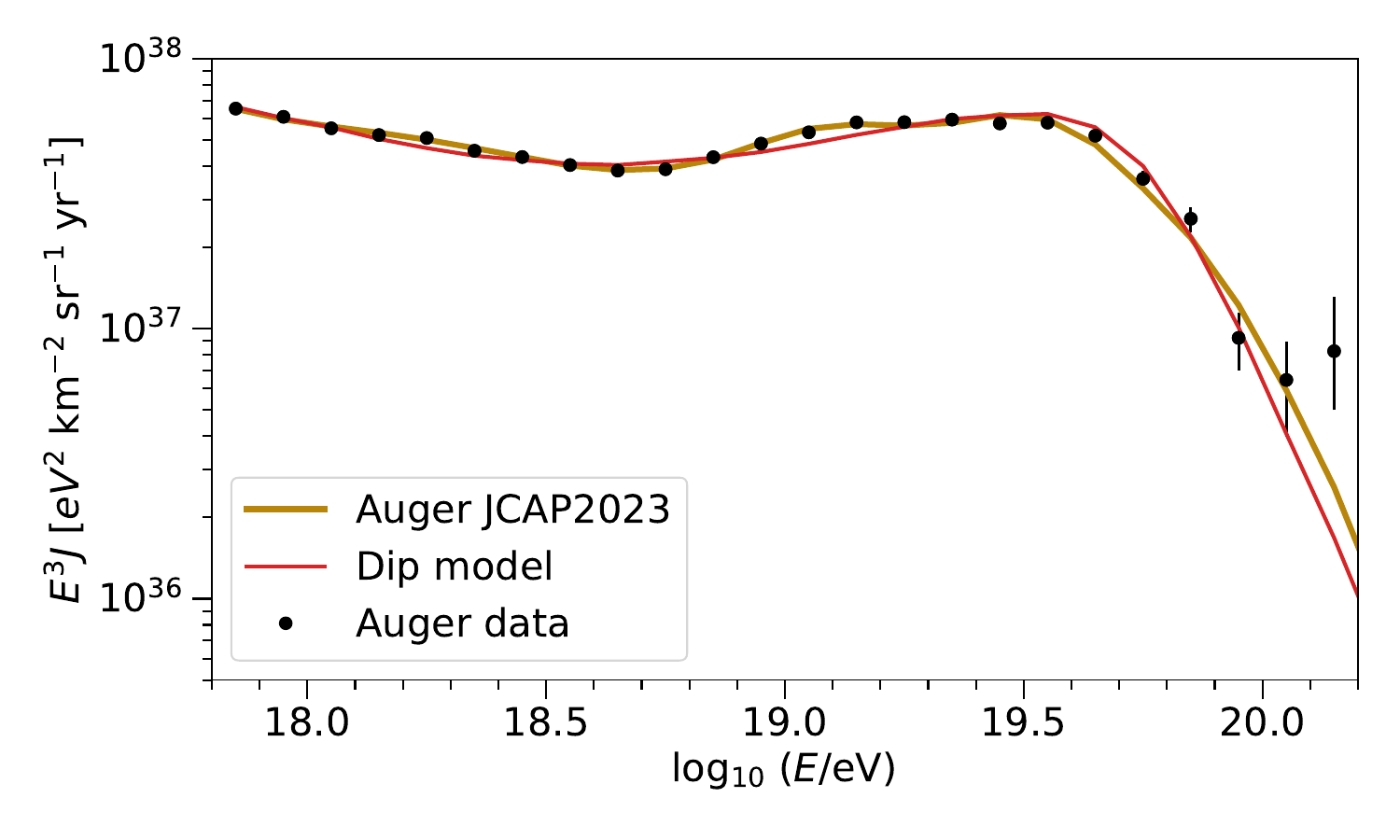}
\end{minipage}
\vspace{-0.5cm}
\caption{Left: Intensity of cosmic rays, multiplied by $E^3$ estimated using five different techniques~\cite{PierreAuger:2021ibw}. Right: The SD energy spectrum, multiplied by $E^3$, shown together with the best fit of the spectrum and mass composition with the astrophysical model in \cite{PierreAuger:2022atd} and with the best fit of the spectrum with pure protons at injection (spectrum at source: $\propto E^{-\gamma}$ for $E<E_{\mathrm{max}}$, $\propto E^{-\gamma} \exp{(-E/E_{\mathrm{max}})}$ for $E>E_{\mathrm{max}}$; cosmological evolution of the sources modeled as $\propto (1+z)^{m}$, with the best fit parameters being $\gamma=2.25$, $E_{\max}=10^{19.75}$ eV, $m=5$).
}
\label{spectrum}
\end{figure}
In this section we review the UHECR energy spectrum, mass composition and arrival directions measured at the Pierre Auger Observatory, and describe the current understanding of the characteristics of UHECRs. 

The energy spectrum of UHECRs has been measured with different techniques across the ultra-high-energy range, as collectively reported in Fig.~\ref{spectrum} (left), where the spectra of the 1500 m array using vertical events, inclined events, hybrid events, events detected by the 750 m array and the FD events dominated by Cherenkov light are shown~\cite{PierreAuger:2021ibw}. The events have been scrutinized as a function of their declination, with no evidence for a declination dependence of the spectrum~\cite{PierreAuger:2020qqz}.
Several deviations from a constant power law across the energy are measured in the spectrum at Earth~\cite{PierreAuger:2020qqz}. These might reveal imprints of the properties of UHECR acceleration, the distribution of sources in the local universe, and the effects of the propagation of UHECRs through background photons~\cite{PierreAuger:2020kuy,PierreAuger:2022atd,PierreAuger:2023htc}. Attempts to reproduce the UHECR spectrum with a basic astrophysical model fail if only UHECR protons are assumed to escape from astrophysical accelerators (as in the "dip model"~\cite{Berezinsky:2002nc}). An example is reported in the right panel of Fig.~\ref{spectrum}, where the best fit in terms of propagated protons with a power-law spectrum at the source is shown together with an alternative model, which allows heavier nuclear species to be emitted from sources, as in~\cite{PierreAuger:2022atd}, constrained thanks to the spectrum and mass composition data. In addition, the astrophysical parameters to get the best fit with UHECR protons correspond to those which maximize the expected cosmogenic neutrinos (i.e. high cosmological evolution, $m=5$), which are already excluded by the most up-to-date upper limits~\cite{PierreAuger:2023pjg,PierreAuger:2023mid}. Therefore, even without taking into account the mass composition data, the current precision of the UHECR energy spectrum measurements could be used to exclude the dip model. In fact, the fine structure of the energy spectrum is not reproduced if only propagated protons are taken into account, especially because of the shape of the "ankle" at about $10^{18.7}$ eV and the recently observed "instep" at 13 EeV, and for the implications on the expected neutrino flux.

\begin{figure}[t]
\centering
\includegraphics[scale=0.45]{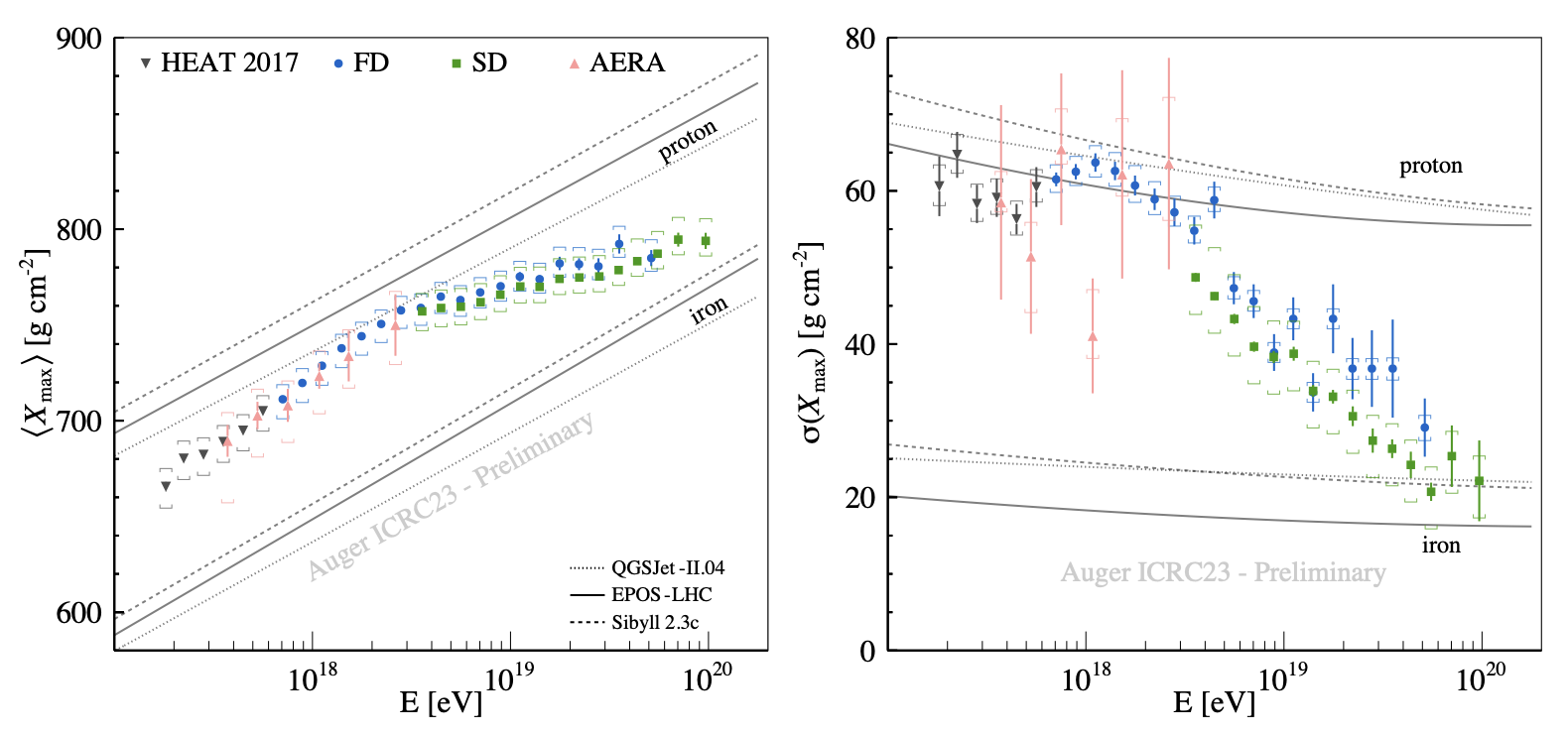}
\vspace{-0.5cm}
\caption{The first (left) and second (right) moments of $\langle X_{\mathrm{max}} \rangle$ distributions as a function of the energy as measured with the FD, the SD, AERA, and HEAT \cite{PierreAuger:2023bfx}.}
\label{Xmax}
\end{figure}
The mass composition of UHECRs can be inferred by the distributions of the depth in the atmosphere at which the shower reaches its maximum number of particles, $X_{\mathrm{max}}$, as measured by the FD. The first and second moments of this distribution reveal the mass composition if compared to the expectations from hadronic interaction models tuned on LHC measurements, that differ in a range of about 20 $\mathrm{g/cm^{2}}$. It is possible to pinpoint an increase of the $\langle X_{\mathrm{max}} \rangle$ as a function of the energy (Fig.~\ref{Xmax}, left), and the observed change in the elongation rate argues for the average composition becoming lighter up to $10^{18.4}$ eV and heavier afterwards. The second moment, $\sigma(X_{\mathrm{max}})$ shows a trend
towards heavier and less mixed composition above $10^{18.6}$ eV~\cite{PierreAuger:2023bfx} (Fig.~\ref{Xmax}, right). Thanks to the use of SD data and a deep-learning-based reconstruction algorithms
~\cite{PierreAuger:2024nzw,PierreAuger:2024flk} it is possible to increase by 10-times the statistics with respect to FD measurements, confirming the evolution of $\langle X_{\mathrm{max}} \rangle$ with energy and in particular excluding the absence of breaks at a level of 4.6$\sigma$ and featuring three breaks that are observed in proximity to the features in the energy spectrum. The analysis of the correlations between $X_{\mathrm{max}}$ and the ground-particle signals in WCDs can be also considered as a model-independent test of a pure proton-scenario~\cite{PierreAuger:2024neu}.
In addition, converting the first two moments of $\langle X_{\mathrm{max}} \rangle$ to $\langle \ln A \rangle$ and $\sigma^2(\ln A)$, it appears that using QGSJetII-04 the $\sigma^2(\ln A)$ is negative, therefore showing that this model is not recommended to be used for inferring the mass composition of UHECRs. 

\begin{figure}[t]
\hspace{-0.5cm}
\begin{minipage}{5cm}
\includegraphics[scale=0.4]{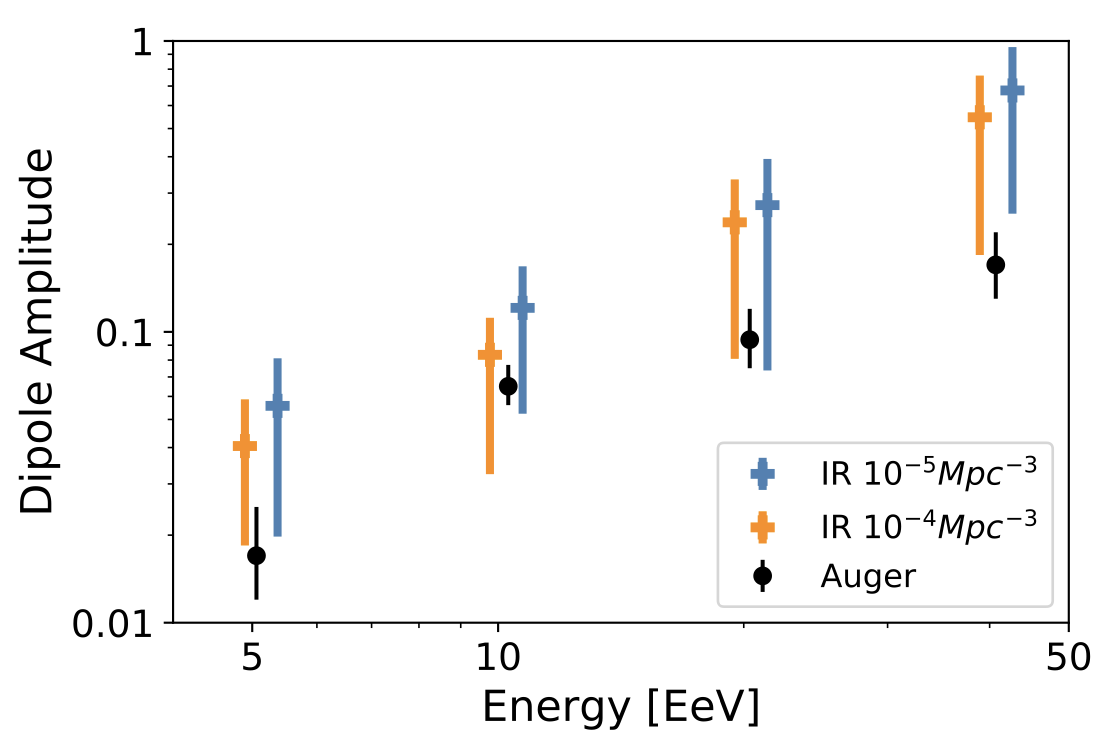}
\end{minipage}
\hspace{3.2cm}
\begin{minipage}{5cm}
\includegraphics[scale=0.19]{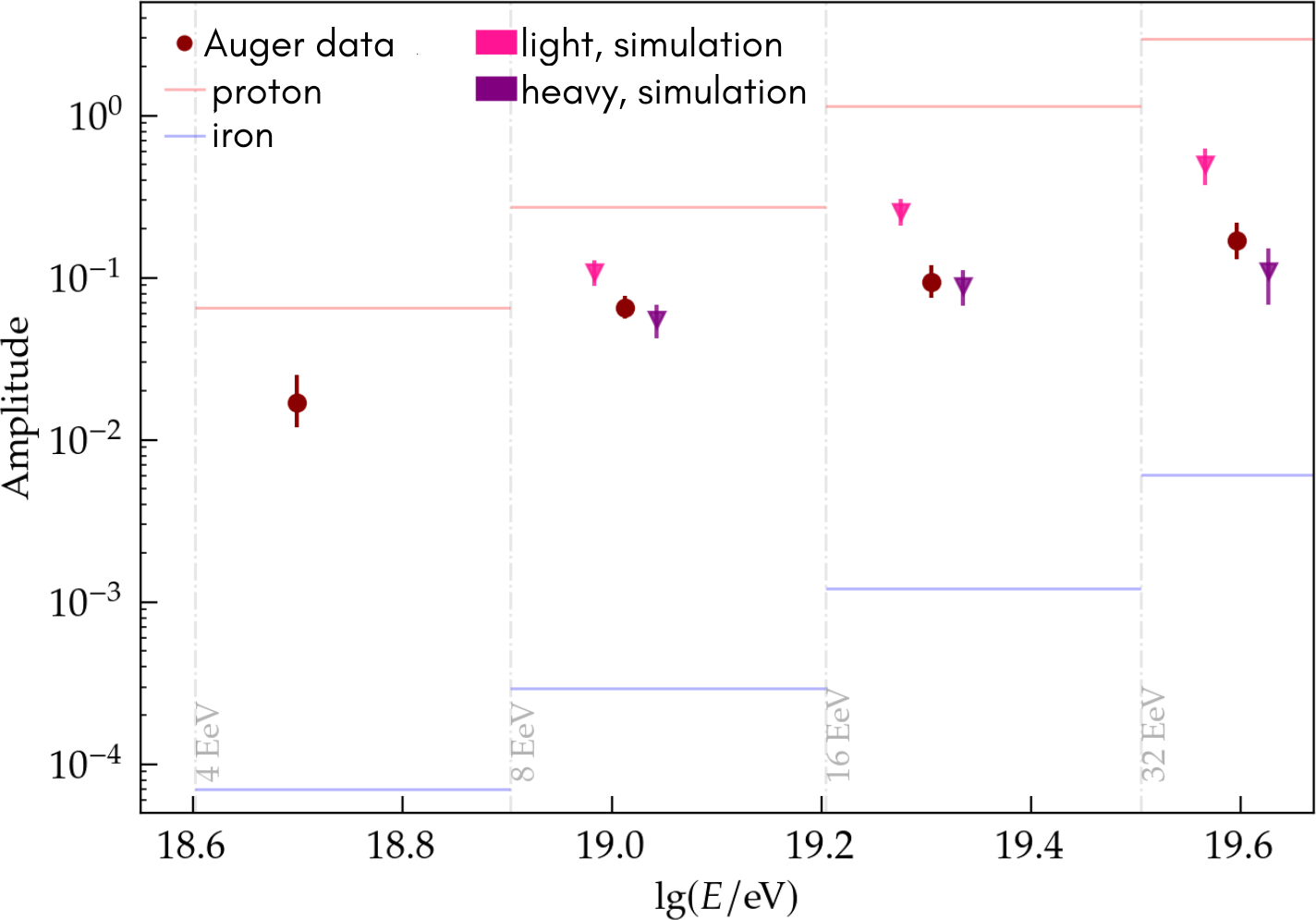}
\end{minipage}
\vspace{-0.4cm}
\caption{Left: Median and 68\% CL range of the dipole amplitudes for the two source densities considered, $10^{-4}$ $\mathrm{Mpc^{-3}}$ (orange) and $10^{-5}$ $\mathrm{Mpc^{-3}}$ (blue); data points are shown with black dots~\cite{PierreAuger:2024fgl}. The four energy ranges are (4-8, 8-16, 16-32, $\geq$ 32) EeV. Right: Expected dipole amplitude in mass-selected subsets of data (probed on simulations).}
\label{dipole}
\end{figure}

Several searches for small- and intermediate-scale anisotropies are carried on at the Auger Observatory, including searches for localized excesses, autocorrelation and correlation with structures~\cite{PierreAuger:2022axr}. The correlation of UHECR arrival directions with the flux pattern expected from catalogs is evaluated against isotropy using likelihood-ratio analyses, and the highest significance is found for starburst galaxies. Both the catalog-based searches and the search in the Centaurus region point to the most significant signal at an energy threshold close to 40 EeV. 
The strongest signal of the extragalactic origin of UHECRs above 8 EeV comes from the dipole, whose direction points 113$^{\circ}$ away from the Galactic center~\cite{PierreAuger:2024fgl}. The dipole amplitude increases with energy, as possibly due to the larger relative contribution from the nearby sources for increasing energies,
whose distribution is more inhomogeneous, and to the growth of the mean primary mass of the particle with increasing energy. By taking into account astrophysical scenarios obtained
from the interpretation of the energy spectrum and mass composition, the dipole amplitude is shown to be compatible with the
matter distribution of the large-scale structure~\cite{PierreAuger:2024fgl}, as reported in Fig.~\ref{dipole}, left. In addition, as shown in the right panel of Fig.~\ref{dipole}, analyses probed on simulations show that by defining light and heavy populations through a mass
estimator with universality, we will possibly observe a separation in the total dipole amplitude in mass-selected subsets of data~\cite{Emily:UHECR24}.

We can summarize at this point that UHECR data show features increasingly significant and independent of models, such as the following ones: (i) several changes of the spectral index of the measured energy spectrum are firmly established, including the recently observed "instep", (ii) the absence of breaks in $\langle X_{\mathrm{max}} \rangle$ can be rejected with high significance, and (iii) the dipole signal is found to be directed outside the Galactic center, and its amplitude increases with energy.
These experimental evidences can be used to challenge basic astrophysical scenarios such as the UHECR proton-paradigm, which was the leading scenario to explain UHECRs at the time of the birth of the Pierre Auger Observatory. The astrophysical picture emerging from these measurements was not expected at that time. 
It is interesting to analyze the behavior of the $X_{\mathrm{max}}$ moments as a function of the energy to get insights not only on the mass composition, but more generically on UHECR astrophysical scenarios. In particular, the $\sigma(X_{\mathrm{max}})$ as a function of the energy comprises the information of both intrinsic shower-to-shower fluctuations and the logarithmic dispersion of the masses as they hit the Earth atmosphere, which are due to the spread of nuclear masses at the sources and to the modifications occurring in their propagation to Earth~\cite{PierreAuger:2013xim}. The larger is the spread among the nuclear species, the broader is the combined distribution of $X_{\mathrm{max}}$~\cite{Kampert:2012mx}; therefore in order to reproduce the observed behavior of the $\sigma(X_{\mathrm{max}})$, the superposition of different nuclear species as they hit the atmosphere must be very limited. In terms of astrophysical scenarios, this can be achieved if the spectra of the different nuclear species emitted at the escape of the source environment are almost monochromatic. Such a condition can be satisfied if the UHECR spectra describing the measurements above the "ankle" are power-law functions with hard spectral index and low-rigidity suppression at their sources~\cite{PierreAuger:2016use}, with limited source-to-source variations~\cite{Ehlert:2022jmy}. The observed flux suppression at the highest energies would therefore be ascribed to a combination of propagation effects and of the limited power of acceleration processes at the sources. The hardness of the spectral index of the spectrum at the sources required to describe the spectrum and composition data at the highest energies and the much softer one required below the ankle~\cite{PierreAuger:2022atd,Luce:2022awd} could be explained thanks to the effect of particle confinement in the source environment~\cite{Unger:2015laa}. In-source interactions could also shape the ordering of the mass fractions at the escape of the sources in terms of rigidity and/or of Lorentz factor~\cite{Muzio:2023fng}, depending on their efficiency~\cite{Biehl:2017zlw}, which is then reflected also in the ordering of the mass fractions at Earth~\cite{Trimarelli:2023bnn}.
We stress here that the hardness of the spectral index found in the astrophysical scenario above the ankle cannot be compared to what predicted for the Fermi acceleration, because the model accounts for the spectra at the escape from the source environment and not at the acceleration; moreover, it can be demonstrated that a magnetic horizon effect could imply a softer spectral index at the sources~\cite{PierreAuger:2024hlp}. While the determination of the amount of protons is relevant to several aspects, including the expectations for cosmogenic particles and the investigation of the efficiency of in-source interactions, on the other side pinpointing the heaviest nuclear species contributing to the UHECR flux at the highest energies will be crucial to scrutinizing the nature of the accelerators. The determination of the mass fractions \cite{PierreAuger:2023bfx,PierreAuger:2023xfc} is therefore a key aspect of the Phase II of the Pierre Auger Observatory for increasing our knowledge about UHECR properties and of their sources.

The properties of the Galactic magnetic field can be investigated with the current and future data of the Pierre Auger Observatory, thanks for instance to the increasing precision of the dipole measurement and to the improvements in the magnetic field modeling~\cite{Unger:2023lob,Bister:2024ocm}. Including the magnetic field effects in analyzing the arrival direction of cosmic rays is essential to test the current evidence of deviation from isotropy of the starburst galaxies~\cite{Higuchi:2022xiv,Luca:UHECR24}, and can possibly allow to constrain the transient nature of UHECR sources~\cite{Marafico:2024qgh}.

Cosmogenic particles are related to the highest energy cosmic rays, being produced in their interactions with background photons. 
The increase of the sensitivity to neutrinos and photons will therefore improve the understanding of the characteristics of UHECR sources such as the ones related to their cosmological distribution~\cite{PierreAuger:2023mid}. 
The sensitivity to cosmogenic particles plays a key role in multimessenger studies such as the point-source searches in spatial and time correlation with other events, such as for instance the binary-neutron-star mergers as studied in \cite{LIGOScientific:2017ync,ANTARES:2017bia}.
Non-standard physics can be investigated thanks to the bounds on cosmogenic neutrinos and photons (Lorentz invariance violation in extragalactic propagation~\cite{PierreAuger:2021tog}, dark matter properties~\cite{PierreAuger:2022jyk,PierreAuger:2023vql}), as well as thanks to the sensitivity to detect upward going showers~\cite{PierreAuger:2023elf,PierreAuger:2023wke} and to the development of showers (Lorentz invariance violation in the development of EAS~\cite{PierreAuger:2021mve}).

At the Pierre Auger Observatory particle interactions can be studied at energies not accessible with terrestrial accelerators. This allows to infer the proton-air and proton-proton cross section~\cite{PierreAuger:2023xfc} where the cosmic-ray composition is dominated by proton primaries. Interestingly, the content of muons measured in the EAS is greater than what is predicted by simulations, as found also in other experimental results. Further insights into the origin of the muon deficit are obtained by looking at the event-by-event fluctuations in the number of muons, which are in good agreement with model predictions~\cite{PierreAuger:2021qsd}. This opens the way to explaining the origin of the muon deficit in the models as possibly due to an effect that accumulates along the shower development. 

\section{Conclusions and future prospects}
\vspace{-0.3cm}
The Pierre Auger Observatory is starting Phase II with detector and electronics upgrades. The data collected in its first 20 years of operation allowed to disfavor scenarios expected to describe UHECRs at the times of its origin, revealing unexpected aspects of UHECR physics as well as insights in particle physics and potential power in constraining physics beyond the standard model. AugerPrime will exploit multi-hybrid techniques, increasing the statistical sensitivity of ongoing searches and testing hypotheses or signatures in data which are not yet significant enough, strongly pushing the knowledge of UHECRs.

\end{document}